# An Optimal Heuristic for Sum of All Prime Numbers – Logic for Large Inputs using RAPTOR

Mr. P. Vasanth Sena
Department of Information Technology
Sree Chaitanya college of Engineering, Karimnagar
vasanthmtechsit521@gmail.com

Mr. Kanegonda Ravi Chythanya
Department of Information Technology
Sree Chaitanya college of Engineering
chythu536@gmail.com

***Abstract*: An optimal heuristic logic is an effective method for finding the sum of all prime numbers up to a given number. This paper presents different approaches, namely, general method and optimal method which facilitate to compare the results and draw the optimal solution. The method adopted is to know the number of symbols evaluated in each logic, construct better approaches based on heuristics, proposals and implementations of human sequential development using Rapid algorithmic prototyping tool for ordered reasoning (RAPTOR). In traditional approach, task is complex in point of time and space; however, this method reduces these prime factors by applying simple mathematical theorems and heuristics. This model effectively works with large numeric inputs. It has been tested on RAPTOR with flow charts, results indicates that algorithms are fast, effective and scalable.**

## I. INTRODUCTION

Raj studying 8th standard in Secundrabad public school. His Math's teacher given a problem that "Find the sum of all prime numbers up to 2M", Where M stands for Million. Assume that you are neighbor of him, how can you help him to solve that problem in optimal way?

Prime number is a number which is divisible by 1 & itself. In other words a number which contains the factors 1 & itself i.e. there are no other factors for that number. The word divisible means suppose if we divide that number by other number then remainder is zero.

In Enhancededu CIT program it is computed that the "SUM OF ALL PRIME NUMBERS UPTO 2M", Where M stands for Million. It is interpreted as 1 Million is equal to 10, 00,000.That need to find the solution for the sum of all prime numbers up to 20,00,0000. In general, all the prime numbers are odd numbers except 2, it is even prime number. So it needed to find the sum of all prime numbers up to 20, 00,000 i.e. 2+3+5+7+11+…

## II. BACKGROUND

In general more real life problems has many solutions these are said to be feasible solutions but in that solutions one solution is more suitable for that problem in terms of time complexity that is said to be optimal solution. In this paper, implemented many logics for that problem and compared these logics in terms of time complexity and number of symbols evaluated. It was concluded that the logic7 was the optimal solution for the given problem.

The solutions presented in this paper were implemented on Rapid Algorithmic Prototyping Tool for Ordered Reasoning (RAPTOR). It gives output as number of symbols evaluated and the program output. It assumed that for each symbol evaluation as one unit of time, and then reduced number of symbols evaluated then accordingly time reduced.

This is simple logic in order to find out the sum of all prime numbers below the given number. By taking numbers from 2, 3,4,….2M, Every time taking one number from the above series in sequential and checking for the factors. If the factors count is 2 then added to sum else skip this number. A positive number is prime number if and only if it is divisible by 1 and by itself only [1]. If it has exactly two factors i.e. count is 2 then it is prime number. In general "Every number is divisible by 1 & itself" [2], so no need to check factors except 1 & itself. If there were others factors rather than above mentioned than it is not prime, otherwise it is prime, and also check count is equal to zero then add to sum. It knows that "Any number n, after n/2 there is no factors for that number except itself" [3], so need to check factors from 2 to floor(n/2), check count is equal to zero then add to sum. "Instead of checking n/2, check for square root of n (sqrt(n))" [4]. So need to check factors from 2 to square root of n, if count is equal to zero then add to sum. "All the prime numbers are odd numbers except 2" [5], hence check for even numbers started from 4 (four) "Odd number is not divisible by even number" [6].

## III. METHODOLOGY

INPUT RANGE**:** The range is 2,3,4…..20,00,000.In general, a division (/) operation gives Coefficient and modulo division (%) operation gives remainder. Example 11/2=5 and 11%2=1. Notice that for each modulo division operation needed one unit of time.

LOGIC1(GENERAL METHOD): This is simple logic in order to find out the sum of all prime numbers below the given number. By taking numbers from 2, 3,4,….2M, Every time taking one number from the above series in sequential and checking for the factors. If the factors count is 2 then added to sum else skip this number.

For e.g., number 11, checking 11 modulo (%) operation with 1 … 11, needed 11 units of time.

11%1 11%2 11%3 11%4 11%5 11%6 11%7 11%8 11%9 11%10 11%11

LOGIC2: In logic1 by making small modifications it becomes better than previous one. In general "Every number is divisible by 1 & itself", so no need to check factors except 1 & itself. If there were others factors rather than above mentioned than it is not prime, otherwise it is prime, and also check count is equal to zero then add to sum.

For e.g., number 11, checking 9 modulo (%) operation with 2 … 10, needed 9 units of time.

11%2 11%3 11%4 11%5 11%6 11%7 11%8 11%9 11%10

LOGIC3**:** By making minor changes in logic2, it becomes effective than previous one. It knows that "Any number n, after n/2 there is no factors for that number except itself", so no need to check factors from 2 to floor(n/2), check count is equal to zero then add to sum.

For e.g., number 11, checking 4 modulo (%) operations with 11%(2..floor(11/2) i.e 5), needed 4 units of time.

11%2 11%3 11%4 11%5

LOGIC4**:** Minor changes to logic 3 gives better than previous one. We know that "Any number n, after n/2 there is no factors for that number except itself", Instead of checking n/2 check for square root of n (sqrt(n)). So need to check factors from 2 to sqrt(n), if count is equal to zero then add to sum.

For E.g., number 11, checking 2 to floor(sqrt(11)) gives 2 modulo (%) operations hence 2 units of time. i.e. 11%2 11%3

LOGIC5(SKIP EVEN NUMBERS IN NUMERATOR)**:** Here, "All the prime numbers are odd numbers except 2", hence check for even numbers started from 4.

Here, we are taking initial value as sum=2, starting number as 3, every time numerator incremented by 2, so we can skip even numbers.

Hence input range is 3,5,7,9,11,13,15,17,19,21,23………19,99,999

LOGIC6(SKIP EVEN NUMBERS IN DENOMINATOR)**:** In logic5, in numerator there is no even numbers, however odd numbers are not divisible by even numbers, in order to improvement in logic remove all even numbers in denominator.

We are taking initial values are sum=2, starting number is 3, every time numerator incremented by 2, so we can skip even numbers.

Hence input range is 3,5,7,9,11,13,15,17,19,21,23………19,99,999

For e.g.,11%2 ,11%3. In these two remove 11%2 then it becomes 11%3

LOGIC7 (SKIP FROM LOOP IF IT IS NOT CONSIDERED PRIME)**:** Modify in above logics we make optimal logic in this stage. In other hand my simple idea is "If number contains one factor except one and itself, no need to check for other factors, it indicates that number is not prime then skips form the loop".

For e.g., Take number 21, According to our previous logics

| | | | | | |
|---|---|---|---|---|---|
| L1: | 21%1 | 21%2 …21%10….. | | 21%20 | 21%21 |
| L2: | 21%2 | 21%3 | 21%4…21%10….. | | 21%20 |
| L3: | | 21%2 | 21%3 | 21%4…21%10….. | |
| 14: | | 21%2 | 21%3 | 21%4 | |
| L5: | | 21%2 | 21%3 | 21%4 | |

L6: 21%3

L7: 21%3

For e.g., take number 45, According to our previous logics

| | | | | | |
|---|---|---|---|---|---|
| L1: | 45%1 | 45%2 | 45%6…45%22…. | | 45%45 |
| L2: | 45%2 | 45%3 | 45%6…45%22….. | | 45%44 |
| L3: | 45%2 | 45%3 | 45%4 | 45%6… | 45%22 |
| L4: | 45%2 | 45%3 | 45%4 | 45%5 | 45%6 |
| L5: | 45%2 | 45%3 | 45%4 | 45%5 | 45%6 |
| L6: | 45%3 | | 45%5 | | |
| L7: | 45%3 | | | | |

**For number 45:**

| Logic | Number of Modulo division operations |
|---|---|
| Logic1 | 45 |
| Logic2 | 43 |
| Logic3 | 21 |
| Logic4 | 5 |
| Logic5 | 5 |
| Logic6 | 2 |
| Logic7 | 1 |

## Flow Charts by RAPTOR:

## Logic1: (General Method)

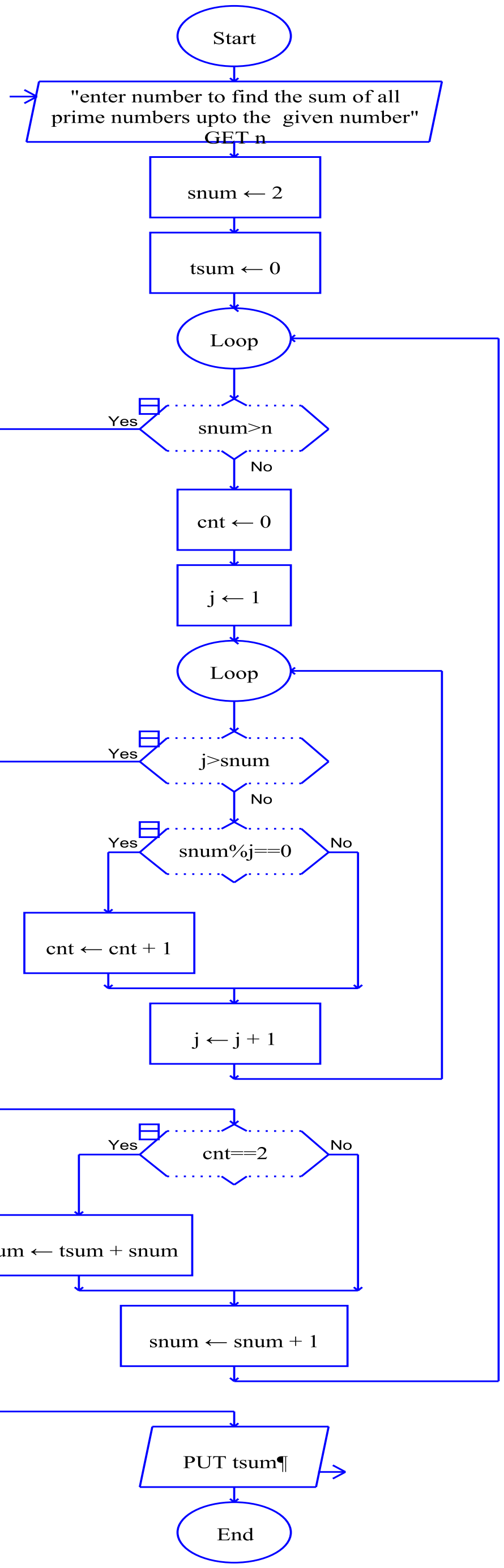

**Logic2:**

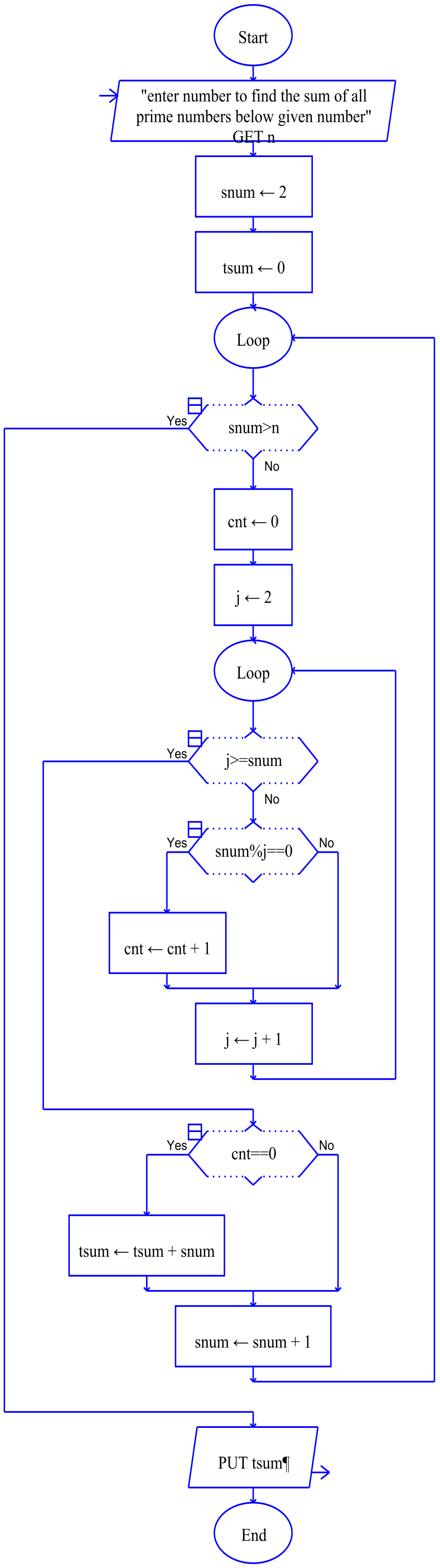

**Logic3:**

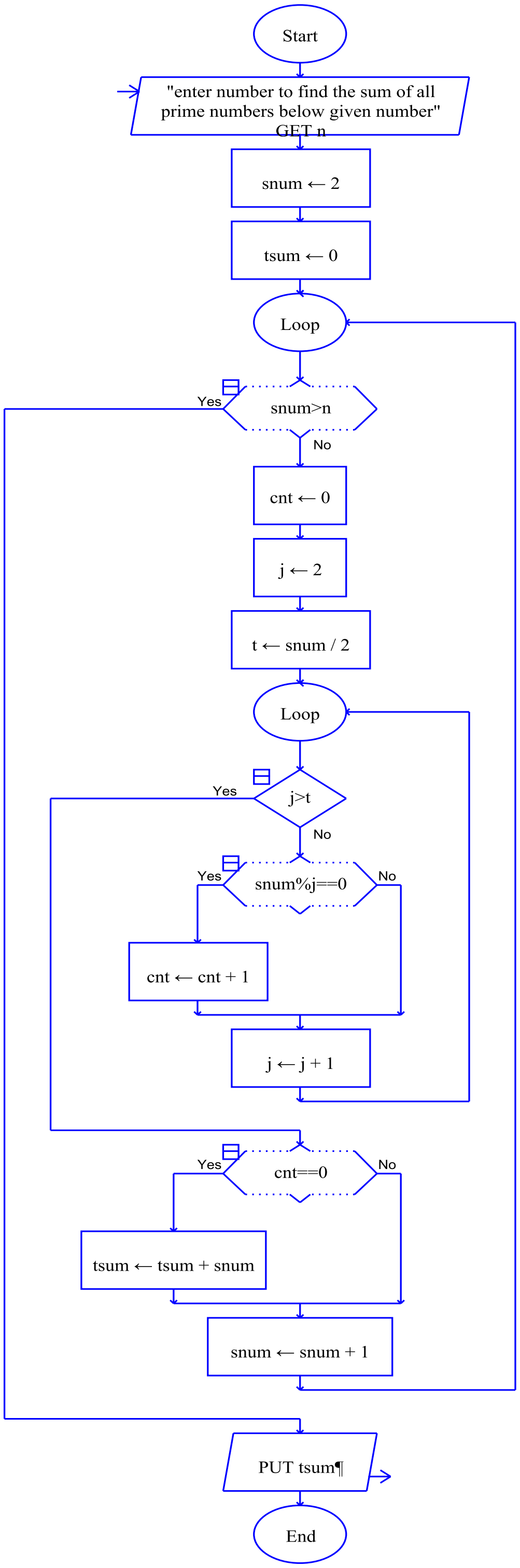

**Logic4:**

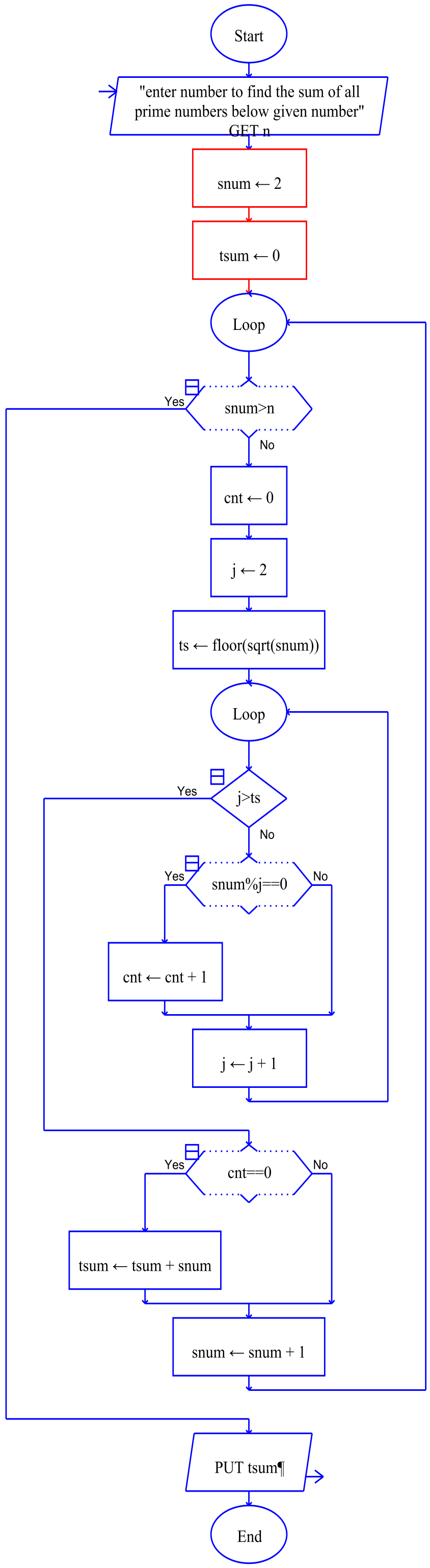

**Logic5:**

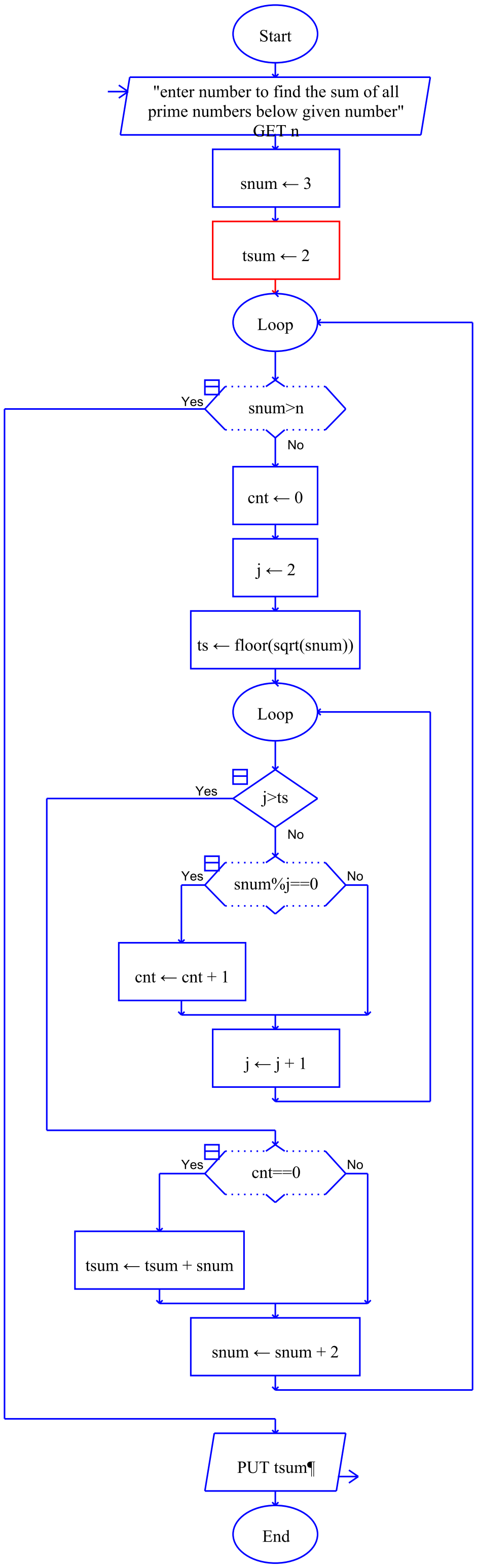

## Logic6:

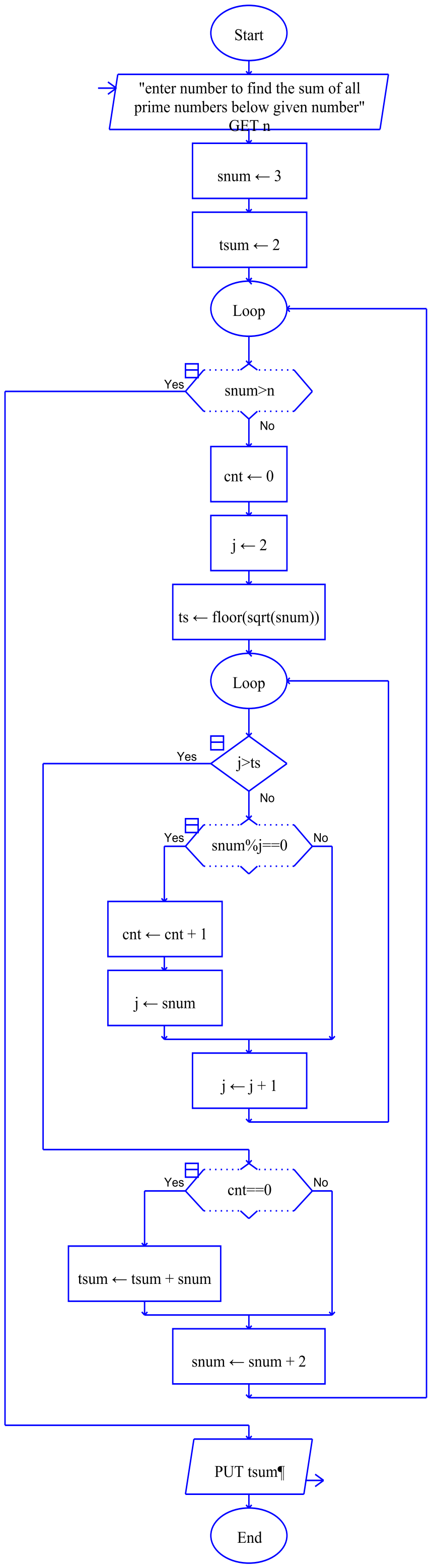

## Logic7:

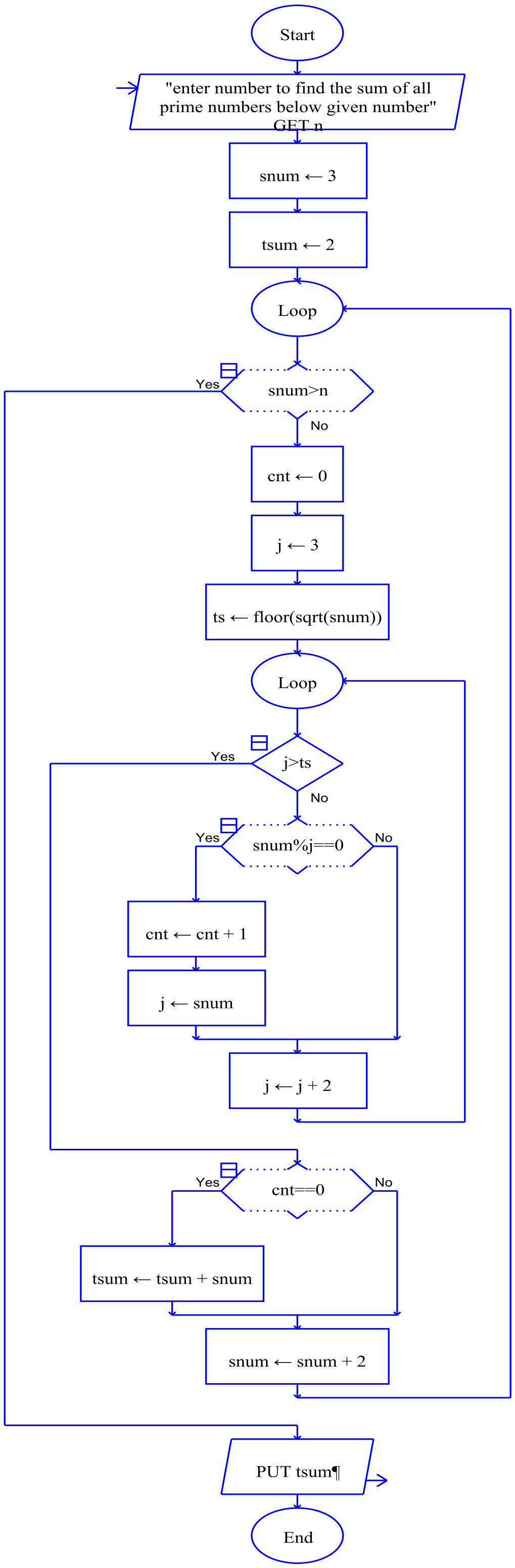

**Algorithm:**

```
Algorithm Optimal_Heuristic_Prime()
{
// Integer variables are n i.e. given number,
//snum i.e. starting number, tsum i.e. total
//sum,cnt i.e. count
Integer n,snum,tsum,i,j,ts,n,cnt;
        tsum :=2;
write "Enter number to find the sum of all
prime numbers up to that number:";
read n
for snum :=3 to n step by 2 do
        {       //begin outer for
        cnt :=0;
        ts:=floor(sqrt(snum));
        for j:=3 to ts step by 2 do
                {       //begin inner for
                if (snum%j==0) then
                 cnt:=cnt+1;
                j:=snum;
                end if
                }       //end inner for
        If(cnt==0) then tsum:=tsum+snum;
        }       //end outer for
        write tsum;
} //end Algorithm Optimal_Heuristic_Prime
```

**Output:**

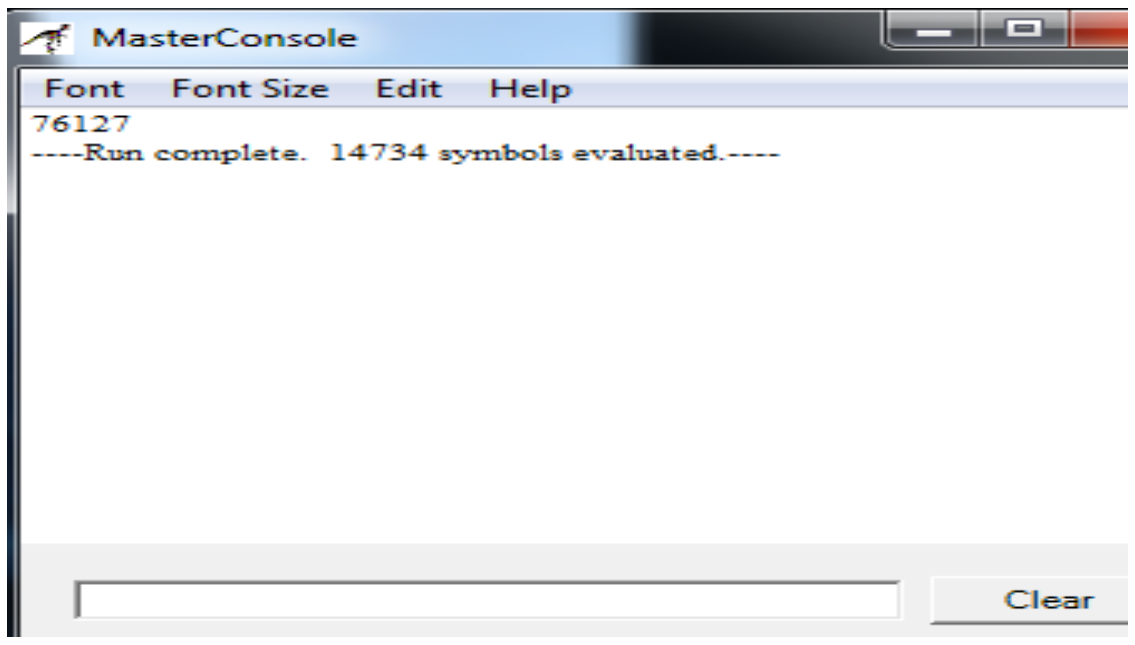

## IV. RESULTS

I have taken input value 1000, the following observations made crucial role in optimal solutions.

| LOGIC | ETDIB | ETNIB | STARTING NUMBER | INTIAL SUM | COUNT CHECKING | #SYMBOLS EVALUATED | OUTPUT |
|---|---|---|---|---|---|---|---|
| L1 | 1 | 1 | 2 | 0 | 2 | 20,17,232 | 76,127 |
| L2 | 1 | 1 | 2 | 0 | 0 | 20,07,242 | 76,127 |
| L3 | 1 | 1 | 2 | 0 | 0 | 10,10,241 | 76,127 |
| L4 | 1 | 1 | 2 | 0 | 0 | 90,177 | 76,127 |
| L5 | 1 | 2 | 3 | 2 | 0 | 44,429 | 76,127 |
| L6 | 1 | 2 | 3 | 2 | 0 | 24,486 | 76,127 |
| L7 | 2 | 2 | 3 | 2 | 0 | 14,734 | 76,127 |

ETDIB: Every time denominator incremented by

ETNIB: Every time numerator incremented by

#SYMBOLS EVALUATED: Number of Symbols evaluated

## V. DISCUSSION

I would like to point out that the algorithm has high computing speed, so it turns the program to quite an optimum state. Furthermore, less coding helps to understand the program easily and you can rewrite it in any programming language.

## VI. CONCLUSION

As I mentioned in the introduction, prime numbers are used in a lot of mathematical calculations, especially, the computations that relate to security that need exact result of prime numbers. So this program can be part of any program that needs prime such as information security algorithms, adaptability in logic development.